\documentclass[preprint,journal]{vgtc}       

\usepackage{mathptmx}
\usepackage{graphicx}
\usepackage{times}
\usepackage{float}
\usepackage{caption}

\usepackage[bookmarks,backref=true,linkcolor=black]{hyperref} 
\hypersetup{
  pdfauthor = {},
  pdftitle = {},
  pdfsubject = {},
  pdfkeywords = {},
  colorlinks=true,
  linkcolor= black,
  citecolor= black,
  pageanchor=true,
  urlcolor = black,
  plainpages = false,
  linktocpage
}

\onlineid{228}

\vgtccategory{System}

\vgtcinsertpkg

\preprinttext{Under review for Scivis 2016}

\title{GraphiteVR: A Collaborative Untethered Virtual Reality Environment for Interactive Social Network Visualization}

\author{Sam Royston, Connor DeFanti, Ken Perlin}

\shortauthortitle{Biv \MakeLowercase{\textit{et al.}}: Global Illumination for Fun and Profit}

\abstract{
The increasing prevalence of Virtual Reality (VR) technologies as a platform for gaming and video playback warrants research into how to best apply the current state of the art to challenges in data visualization. Many current VR systems are non-collaborative, while data analysis and visualization is often a multi-person process. Our goal in this paper is to address the technical and user experience challenges that arise when creating VR environments for collaborative data visualization.  We focus on the integration of multiple tracking systems and the new interaction paradigms that this integration can enable, along with visual design considerations that apply specifically to collaborative network visualization in virtual reality.  We demonstrate a system for collaborative interaction with large 3D layouts of Twitter friend/follow networks. The system is built by combining a ``Holojam'' architecture (multiple GearVR Headsets within an OptiTrack motion capture stage) and Perception Neuron motion suits, to offer an untethered, full-room multi-person visualization experience.
} 


\CCScatlist{ 
 \CCScat{K.6.1}{Management of Computing and Information Systems}%
{Project and People Management}{Life Cycle};
 \CCScat{K.7.m}{The Computing Profession}{Miscellaneous}{Ethics}
}

  \teaser{
 \centering
 \includegraphics[width=16cm]{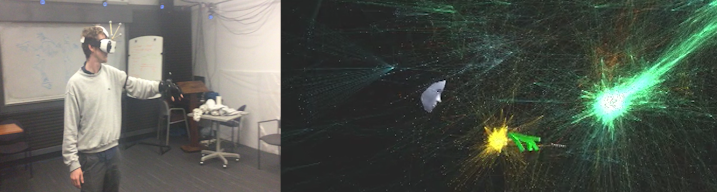}
  \caption{A user of our system as he simultaneously appears in reality and VR}}

\begin{document}


\firstsection{Introduction}

\maketitle


In many applications, the utility of 3D network layouts and visualization is limited by the fact that they must be projected onto a 2D plane. In contrast, Virtual Reality (VR) provides a natural setting for 3D network visualizations by allowing users to interact with complex network structures in a manner similar to how they might interact with real-life structures. Additionally, the current generation of high fidelity gaming-focused VR headsets such as the Oculus Rift, and HTC Vive are tethered and not designed for full-room interaction or real time collaboration among users in the same room. We believe full room usability (sensitivity to large translations) and real time collaboration will be essential components of future VR visualization applications. Correspondingly, the contributions of this paper are twofold:
\begin{enumerate}
    \item We provide a setting designed to render 3D network layouts at scale, with a focus on the aesthetic and UX advantages gained via \textit{full room} VR
    \item We offer the means for multiple users to \textit{collaboratively} inspect complex network structures, interacting with them manually via high fidelity motion capture systems
\end{enumerate}
To test these methods, we work with networks derived from twitter friend/follow relationships, along with metadata about users (nodes) on the graph. We use Holojam \cite{Holojam}, a system that allows for low-latency data streaming to multiple clients in Samsung GearVR headsets in order to create a nomadic VR application for viewing the network graphs. This experience is further enhanced by the use of Perception Neuron gloves to do high-accuracy finger tracking, which allows us to interact with the virtual environment through gesture controls.

Modern social networks, such as Twitter, contain complex and noteworthy structures at multiple scales. For example, a network of interest may include a set of twenty fake twitter accounts managed by one actor, embedded within a larger network of ten thousand users with a global structure influenced by political affiliations. In order to better visualize and appreciate the complexities of such a network, we have developed our system to fully immerse the user at any scale and allow them to use their hands to manipulate network structures. Our data visualization system is unique in that it is within an environment that is nomadic, collaborative, and manipulable with a user's hands and fingers. As a fundamental principle of this work, we believe that one is more likely to study and interact with an object (i.e. an intricate rendering of a network) if those around them can corroborate its existence and interact with it as well.

When the users enter the virtual space, they see a representation of the network graph (Figure 1) along with other users represented as a mask and hand. From there, the user can move and scale the graph in order to appropriately view various graph clusters and highlight different nodes with their finger. All users within the system can see these manipulations. Users can take advantage of these features to analyze and discuss the complex graph structures before them in ways that were previously not possible.
\begin{figure*}[!ht]
\centering
\captionsetup{justification=centering}
\begin{minipage}{.33\textwidth}
\centering
  \includegraphics[width=0.6 \linewidth, angle=90]{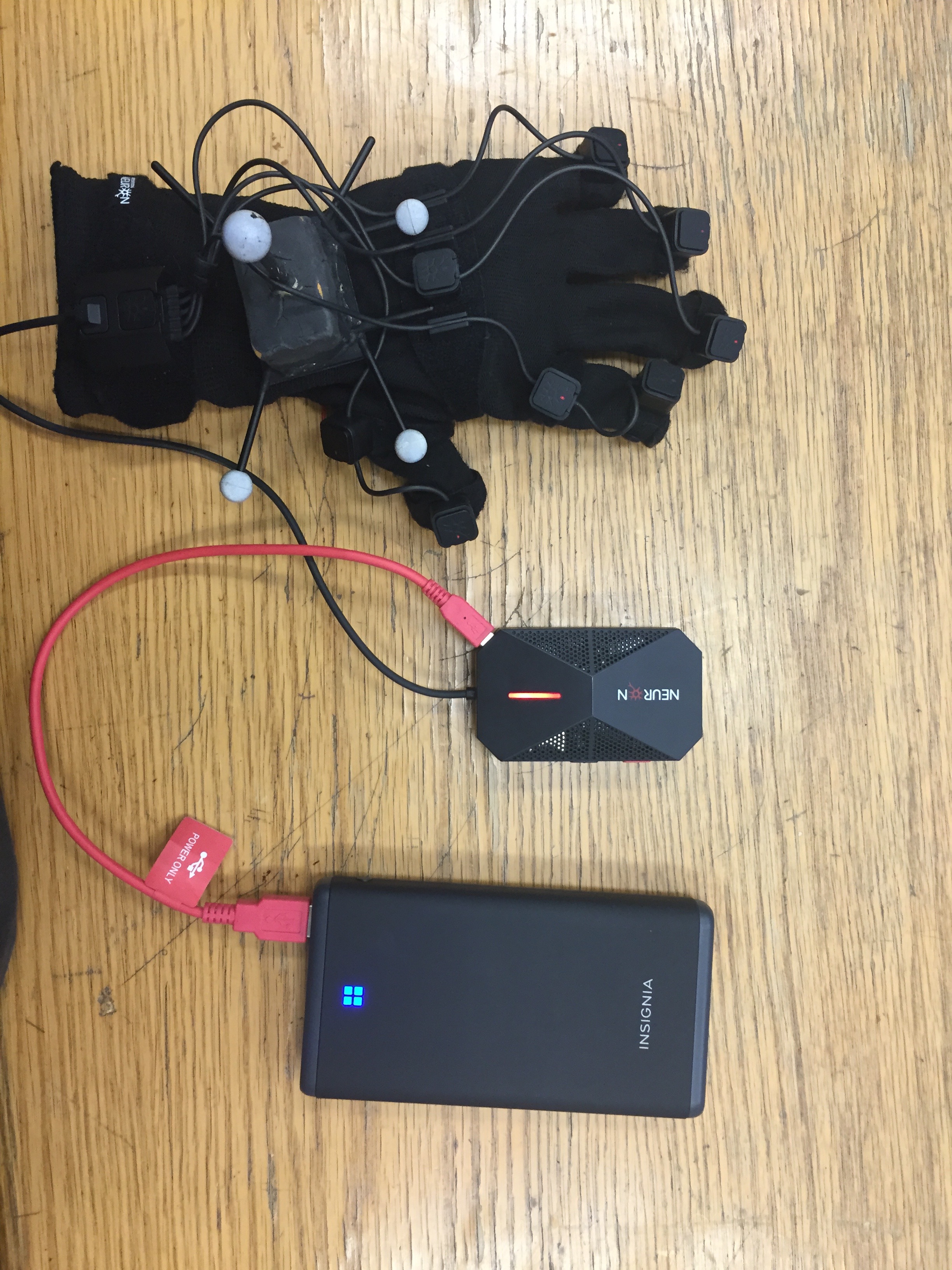}
  \caption{Perception neuron device with additional reflective markers for optical tracking}
  \label{fig:test1}
\end{minipage}%
\begin{minipage}{.33\textwidth}
\centering
  \includegraphics[width=0.9 \linewidth, angle=270]{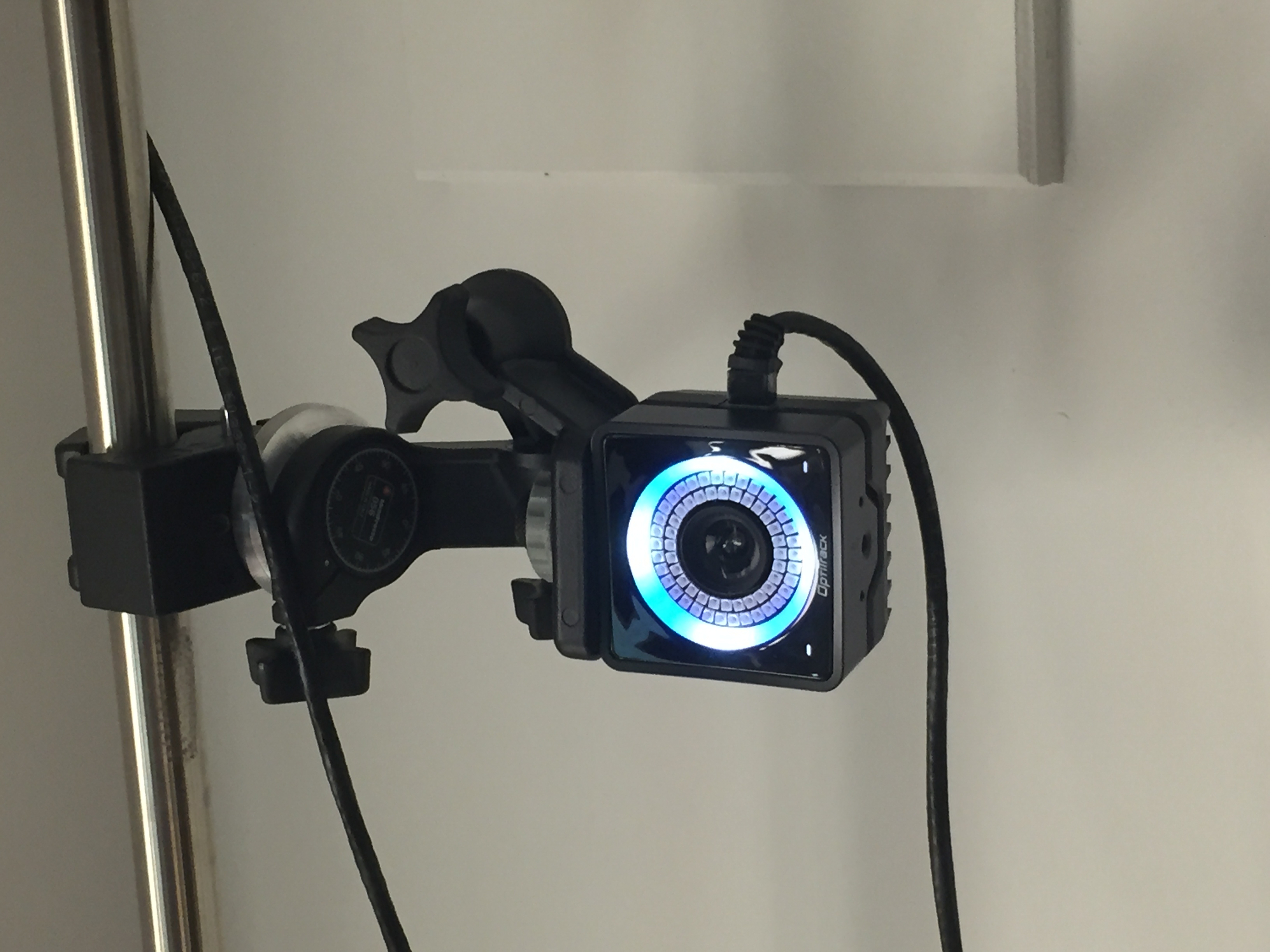}
  \caption{One of the 12 wall mounted Optitrack sensors}
  \label{fig:test2}
\end{minipage}%
\begin{minipage}{.33\textwidth}
\centering
  \includegraphics[width=0.7 \linewidth]{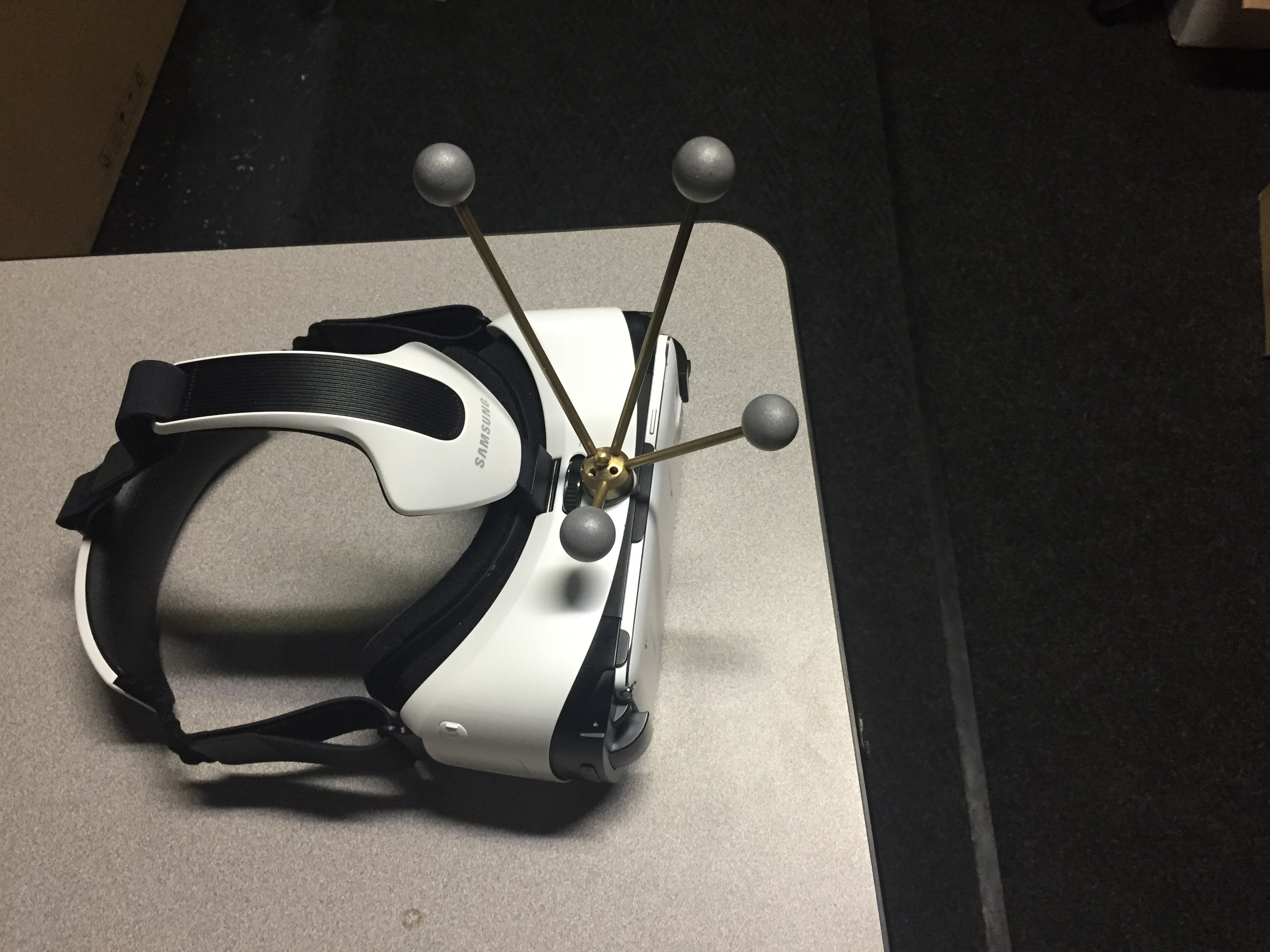}
  \caption{GearVR headset with \\reflective markers}
  \label{fig:test2}
\end{minipage}
\end{figure*}
\section{Related Work}
There have been many decades of research and development into VR technology, both for hardware and software. VR systems have existed as early as the 1960's with Ivan Sutherland's ``Sword of Damocles'' \cite{sutherland1968head} and have greatly expanded since then. Interactive, multiperson systems, such as the CAVE \cite{Cruz-Neira:1992:CAV:129888.129892}, have broadened the applications for VR to be more collaborative and physical, like what our system aims to do, but has the downside of restricting the rendering to the viewpoint of one user. Since then, personal head-mounted displays (HMDs) have become a lightweight, commodity product, such as the Oculus Rift or the HTC/Valve Vive system. These allow for a user to get a high-quality, accurate depiction of their viewport in a virtual world. However, these systems fall short for our purposes, as they are tethered into a typically large computer. Instead, we aim to use the Holojam architecture, discussed in a later section, which allows multiple users to walk around in the space without having to worry about tangled cables. Thus, while other current systems have allowed for data visualization in a 3D virtual environment, ours is the first that allows for collaborative, HMD-based data visualization.

The Gephi and Cytoscape desktop applications \cite{gephi,cytoscape} are some of the most widely used software for visualizing network topologies. These tools provide extensive functionality in terms of layout algorithms, clustering methods, styling and more. Our work can be viewed as a VR front-end to analytics tools like these. In a previous version, the system accepted \texttt{.gexf} files exported from Gephi as input, but using gephi as the core analysis engine proved to be too constrining in terms of metadata annotation and analysis automation. In our current system, the input is a pickled python-igraph \texttt{Graph} object and the analytics and layout is delegated to an auxiliary server utilizing the igraph library (see figure 5).

Work by Donalek et. al. \cite{DBLP:journals/corr/DonalekDDCWLNZLYMGD14} uses uses Unity3D and Oculus VR headsets to visualize astronomical datasets, but has key differences in terms of capabilities and constraints. The rendering in \cite{DBLP:journals/corr/DonalekDDCWLNZLYMGD14} occurs on a personal computer as opposed to a mobile device, and therefore is less resource bound than in our context. Furthermore, the underlying system that we build upon is designed \textit{expressly} for collaborative full-room VR, allowing users to interact with a social network in much the same way they would a real object.

With \textit{Vister}, Heer et. al.\cite{heer2005vizster} provided a precedent for social network visualization using force layouts. We provide much of the same basic functionality in VR, but at a much larger scale. Although not the focus of this paper, a Three.js/webGL frontend that in many respects resembles a 3D version of Vister is additionally provided by the analysis server (discussed below). One contrast between our work and Vister is that the latter focuses on active layouts, while our layouts are computed ahead of time, due to the larger network size. 

The work of Munzner \cite{munzner} offers a comprehensive exposition of large network visualization, covering many types of layout techniques. 

\section{System Architecture}

The architecture described below is largely based on the architecture of the NYU Holojam system, an untethered, multi-user VR system presented at the SIGGRAPH VR Village 2015 \cite{Holojam}. While this system is not the contribution of the paper, we present a brief outline of the hardware and network protocol specifications, as they are important to discuss the primary features of the paper. We have adopted these specifications from Holojam, which we briefly describe below. On top of this system, we have added additional features, such as integration with the Noitom Perception Neuron \cite{https://neuronmocap.com/} glove for hand pose based interactions with the virtual environment and a server for analyzing and distributing graph data from external Internet sources, such as Twitter, for data visualization.
\subsection{Hardware}
The system uses the Samsung GearVR \cite{gearvr}, a lightweight headset that contains a 1000Hz refresh rate inertial motion unit (IMU) to report the change in user orientation to the headset, with rendering powered by a Samsung phone. We chose to use the developer version of the GearVR along with the Samsung Galaxy Note 4, as it offered the largest screen size at the time of our experiments.

The GearVR offers smooth head orientation tracking, but it lacks certain capabilities required for an untethered experience. Primarily, because the GearVR is not inherently built for a nomadic experience, it lacks ground truth in both positional and rotational tracking. Furthermore, it does not even include any form of positional tracking. This means that without an external positional tracking system, the user will not know where they are in space. Additionally, without the ground truth, the user may be facing a different direction in the virtual space than the physical space. If two or more users are in the space, they could potentially be viewing each other in the wrong location without a ground truth.

These issues can be resolved by introducing some form of ground truth tracking into the system. Many such tracking solutions exist today. Holojam works using optical motion capture technology, which can provide the highest quality data and has the advantage of being relatively portable and easy to set up the system. OptiTrack Motive, a well-known motion capture system, allows for 6 degree of freedom (6DOF) tracking at up to 240 frames per second with under 10ms of latency \cite{optitrack}. This information is broadcast using the wireless protocol to each of our GearVR clients.

Using this motion capture software for tracking has its advantages and disadvantages. The advantages of using OptiTrack is that it is fairly reliable and very fast, which is extremely important for delivering timely data to the headsets. The biggest downside to motion capture technology is its cost. High quality motion capture cameras are widely used in academia and industry but are well outside of the consumer price range. On top of this, motion capture is limited by visibility and physical interference. If a marker is obscured, the user loses all positional data until the marker set is visible again. Loss of frames can often cause motion sickness for the user, even if the user loses tracking for as little as half a second. Ultimately, the benefit of having quick, accurate data outweighs the disadvantage of the costs for research purposes, and the visibility issues are easy to avoid with careful camera and marker configurations.

In addition to the optical motion tracking discussed above, we also use sensor-based tracking through the Noitom Perception Neuron system in order to track precise finger motions. These motions allow us to manipulate and interact with graphs with intuitive and natural gestures. This will be discussed further in a later section. The Perception Neuron can transmit data over WiFi, and thus matches our system requirement of being untethered from a central computer. It does, however, require an external power source, but this can be a relatively light battery pack, and so it does not overburden the user.

Since the Perception Neuron system is based entirely on IMUs, it is prone to drift and it does not have a ground truth location. As with the GearVR, we resolve this issue by adding an optical tracker to the gloves, allowing our OptiTrack system to position the hand. From there, the Perception Neuron system positions and orients the fingers relative to the location reported by the OptiTrack.
 \begin{figure*}[!ht]
     \centering
     \includegraphics[width=15cm]{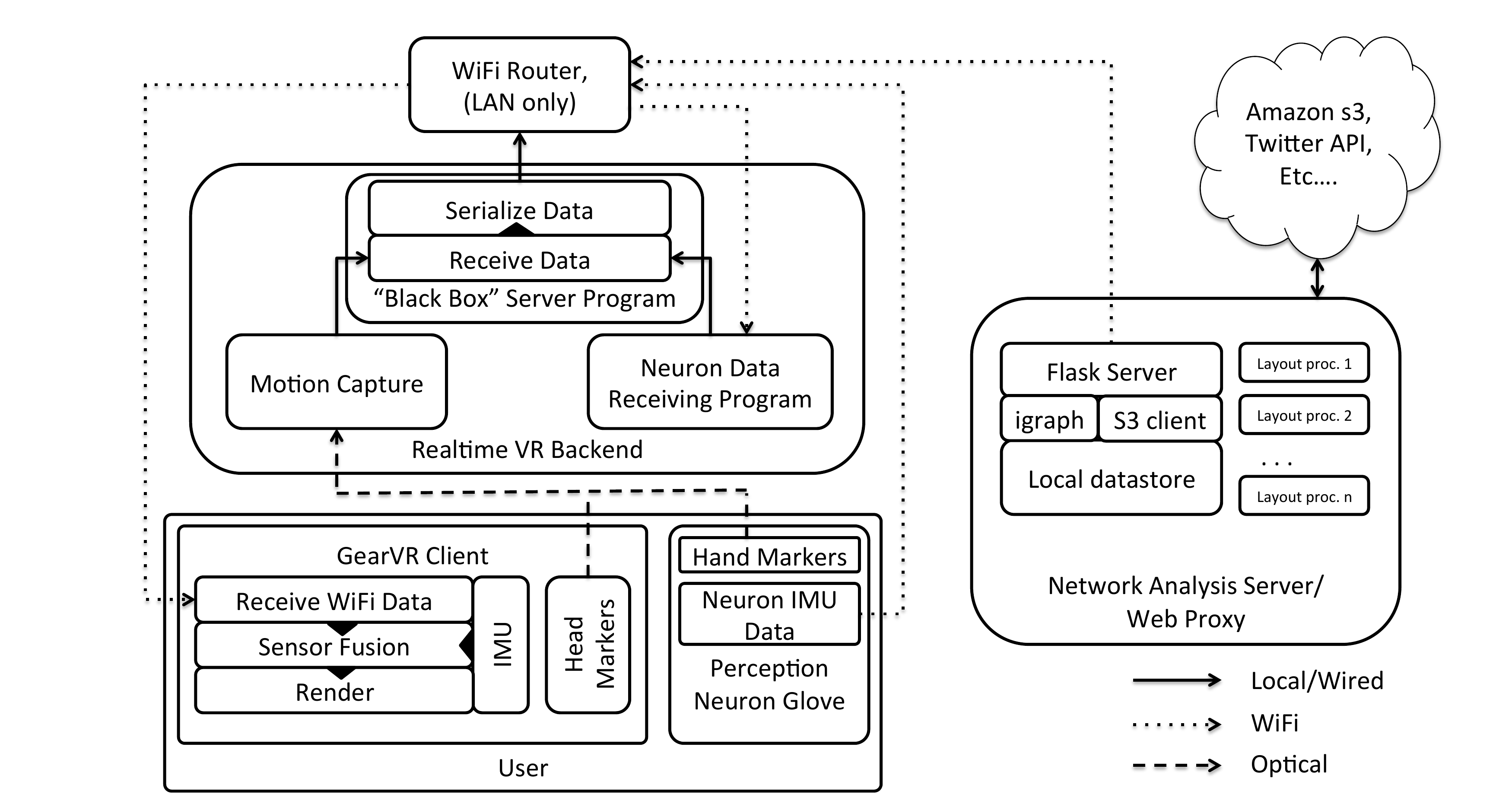}
     \caption{A model of the communications within our system. Upon application launch, data is collected from the Network Analysis Server, which has network graph layouts of several datasets from sources such as Twitter. Once the program is running, there is a constant motion-to-photon loop. This loop starts with the user movement, which is tracked through the OptiTrack and Perception Neuron devices. The Neuron data is sent via WiFi to a receiving program, where it is then processed by the Noitom Axis software. The optical motion capture data is processed by the OptiTrack Motive software. Both forms of motion capture data are sent locally to a main server, the "BlackBox Server," where it is packaged into the Google Protobuf format, and sent over a WiFi UDP multicast stream to each of the phone clients, where it is combined with the GearVR sensor data and rendered on the Unity client.}
     \label{fig:network-layout}
 \end{figure*}
\subsection{Low-Latency Network Protocol}
Figure \ref{fig:network-layout} also briefly summarizes the combined schema of the low-latency network protocol and the Network Analysis Server discussed in the next section.

In order to service tracking data to each of the mobile phone clients in a timely fashion, Holojam uses a lightweight protocol that emphasizes rapid delivery over guaranteed delivery. While this has its fallbacks, which will be discussed later, it allows us to transmit the majority of our data with low enough latency to be imperceptible to the user. In order to achieve the above, it uses a modified UDP multicast protocol in order to transmit the tracking data from the central server to the phone. Note that this protocol is not used to transmit the Twitter data to be visualized, as that data does not need to arrive at the same rate as the motion capture data. That transmission will be discussed in a later section.

The Holojam protocol uses Google Protobuf as a data format, as it is much more compact than many other common data protocols, such as XML \cite{protocolbuffers}. As we mentioned above, since we care more about low-latency delivery over reliability, it is better to use UDP over TCP and multicast over unicast, as both of those methods avoid acknowledgements that can slow down transmission rates. Furthermore, Holojam uses a router with a modified firmware in order to avoid limitations that are set in place for general good practice, but would impede our software, such as multicast rate limiting.

One other issue the server software avoids is IP fragmentation. Typically, WiFi packets will be fragmented into several chunks according to the specifications in the router, and are reassembled on the receiving end. If a number of chunks are missing, then the entire packet will be discarded. Since this can lead to undesirable frame drops, the server instead creates packets that are lower than the router’s maximum transmission size. If the data we wish to send exceeds this size, such as if many more users are present in the area and need to be tracked, those packets are manually broken up into small enough packets, and each one is sent individually.

Finally, we have data packets coming from several different sources. In our case, these different types were the OptiTrack Motive software and the Noitom Axis Neuron software. We needed to ensure that each packet type would be sent fairly. That is, even if one packet type was being received by the server to forward to the phone clients in large volume, the other packet type must be sent through as well. Failing to do so resulted in loss of real-time data, which produced unacceptable latencies. To correct this, we modified the architecture by implementing a simple procedure to rotate through packet types sent, prioritizing packet types that had been sent the least recently. In doing this, we ensured that all packets could be transmitted at even rates.

\subsection{Network Analysis Server}
\subsubsection{System Design}

The network analysis server exists on a different machine from the above system, connected to the optimized lossy WiFi LAN explained above as well as a wired internet connection. New layout requests, based on a pickled (a python serialization protocol) igraph object are fielded via the browser-based web frontend of the analysis server. The server then forks a layout process which will independently operate until completion, at which point the result is uploaded to remote storage service. The state of the layout and upload process is catalogued in a local Mongo instance.  
Once a layout process is finished, a JSON version of the network data is saved in amazon s3 storage, annotated with vertex position data along with any other network analyses metadata performed by the analysis server during processing. Since the analysis server is connected to the same LAN, the headset device(s) running unity can then make requests directly to the local address of the analysis server to gain access to the completed json chunks. Upon downloading, the headset device must parse the json string (usually between 3-10MB) and then configure the appropriate Unity objects. We found that on the Galaxy Note headset devices the string parsing often took much longer than the download itself. For rendering the network objects, we chose to use the \texttt{Mesh()} object attatched to a \texttt{GameObject} in Unity. Using individual GameObjects for each vertex would be convenient because of all the built in functionality they provide, but this approach is untenable for larger network topologies with thousands or tens of thousands of nodes.

\subsubsection{Layout and analysis}

In order to compute layouts (prior to visualization), we use igraph’s 3D Fruchterman-Reingold layout function \texttt{graph.layout\_fruchterman\_reingold\_3d()} using the default cooling exponent of $1.5$ and maximum of $2000$ iterations. Fructerman Reingold takes $O(\left\vert N \right\vert ^2  + \left\vert E \right\vert)$ time per annealing step, and large network structures often require many iterations to produce a satisfactory layout. For this reason we opted to fork separate long-running layout subprocess for each layout task.

The coloration of the nodes corresponds to clusters determined by the  modularity maximization algorithm described in \cite{modularity} and is intended to aid in identifying the different subgroups within the graph. The colors determined by modularity maximization almost always coincide with the agglomerations which are visible within the layout. 

This research was in part intended to test the rendering limits on the GearVR, and once those limits have be reached the experience can become highly unfavorable for the user; low framerates have even been known to induce nausea in some VR users. One strategy to deal with the rendering problem for large graphs is down-sampling the network structure. The Network analysis server provides four options for network sampling:
\begin{description}
    \item[Random Node (RN): ]{Each vertex is included with probability $p$. This scheme is the simplest and yields networks that maintain the degree distribution of the original network relatively well.}
    \item[Random Edge (RE): ]{Each edge is included with probability $p$ and only the nodes they connect are included in the down-sampled version (no singletons). While this scheme directly addresses the rendering issues caused by too many edges, it drastically changes the degree distribution \cite{largegraphs}. We also found that in our experience, RE sampling yields a post-layout spatial structure that is visually very different from the one of the network it is derived from.}
    \item[Random Walk (RW): ]{Begin a random walk, selecting the next node from the current set of neighbors uniformly. In order to prevent the walk from getting 'stuck' in only one area of the network, with probability $p$ transfer to a new random node. We continue the process until a certain proportion of nodes are visited. When $p = 1$ this method is nearly identical to RN sampling. If we select the correct $p$, this method can yield down-sampled graphs with very similar degree distributions. One issue is that $p$ may depend on particulars of the network topology at hand.}
\end{description}

More graph sampling techniques with respect to the goal of preserving degree distribution are discussing at length in \cite{largegraphs}.

\section{VR Interaction Design}

In order to take full advantage of our collaborative VR data visualization system, we designed a few ways for a user to interact with each other and the network graphs. We use the Perception Neuron data to make a simple recreation of each user’s hand, calculate the gesture from that recreation, and then use the gesture to control interactions. Currently, we have implemented gestures to allow the moving and scaling of network graphs.

\subsection{Hand Pose Recognition}

The Perception Neuron data, as noted above, uses a WiFi interface to connect to a server computer that runs the Noitom software used to interpret the Neuron data. This reconstructs a skeleton of the user based on which Neurons are used. In our case, we only use the single-arm model, so we get data for each user’s left arm, including finger movements. Since the program does a full skeletal reconstruction, we get data for several joints, including one for the upper arm, one for the lower arm, one for the hand, and four for each finger. However, we found that we can consistently infer poses with a subset of these data, so to minimize the amount of streaming data, we reduce the hand model to nine points per hand: two for each finger, and one for the hand itself.

Once the data is forwarded through the central server and is received on the Unity phone client, we use the nine points to reconstruct the user’s hand using a low-polygon mesh. Then, using the two points for each finger, we determine whether the finger is “open” or “closed” based on the angle between the fingertip and the knuckle. From here, we can get a 5-bit representation of the hand, one bit for each finger, to get 32 possible poses, although we recognize that only a subset of those poses will be comfortable for human use. Nonetheless, this simple yet effective pose recognition opens up many possibilities for controls in VR that avoid bulky and unnatural control schemes.

\subsection{Network Interactions}

 For this particular project, we decided to focus on three interaction types that would allow for simple collaborative interaction with the network graphs. 
 \begin{description}
 \item{\textbf{Grab}}
 
 First, we implemented a gesture that allows a user to grab and move a graph. This simply translates the entire graph in 3-dimensions. We found this to be primarily useful when a group of users would want to explore a portion of the graph that was outside of the bounds of the physical tracking space.
 \item{\textbf{Scale}}
 
 Second, we implemented a gesture that allows a user to re-scale the graph about the point at which the gesture begins. This provides two primary uses: a user can scale the graph down in order to see the entire structure, or a user can scale the graph up to explore dense clusters.   

\item{\textbf{Highlight}}

Grabbing and scaling are \textit{network wide} interactions and therefore can be easily implemented as transformations applied to the network \texttt{GameObject} as a whole. To interact with individual nodes we use a kd-tree to look up which node is closest to the user's index finger. Kd-trees create efficient spatial subdivisions and allow collision testing in $O(\log(n))$ time. Upon initialization, we store relevant meatadata about each node in a hashtable and when a user selects one, the emanating edges are illuminated and text metadata is displayed near the node's position. In our current prototype we simply display the Twitter handle associated with the selected node, but this text could easily be augmented with the other metadata in the hashtable such as: location, description, and profile image.

\end{description}

From here, it would not be difficult to implement many more actions. For instance, individual nodes could be moved, selected, and analyzed. Graphs could be rotated, restructured around certain clusters, and much more. In other tests, we have implemented 3-dimensional drawing, which allows users to highlight and annotate certain portions of the graph.

However, it was necessary to follow to certain design considerations imposed by our server architecture when designing gestures and the corresponding actions. Because all actions are handled on the client with no acknowledgements or data sent back to the main server, our actions were primarily state-based as opposed to event-based. In other words, we avoided having particular events, such as the opening or closing of a hand trigger actions, and instead relied on the state of the hand model. This would sometimes cause issues if a user had received a large amount of packet loss, causing a desynchronization between graph positions or sizes. While these desynchronization problems were infrequent and typically not an issue, it was This issue could be remedied by having a simple master client that shares its state with all of the other clients, and the other clients follow the master client’s state.
\section{Discussion}

We were able to achieve our goal of data visualization in a collaborative VR environment. We found that the ability for users to see reconstructions of other users and their hands greatly enhanced collaborative analysis of complex graph structures. Clusters that would otherwise look overly cluttered due to high edge connections greatly benefited from being distributed across three dimensions. Our primary goal, which we satisfied, was to allow analysts to observe data together in an immersive environment, a tool which was previously difficult or ineffective to do otherwise.

 \begin{figure}[h!]
     \centering
     \includegraphics[width=7cm]{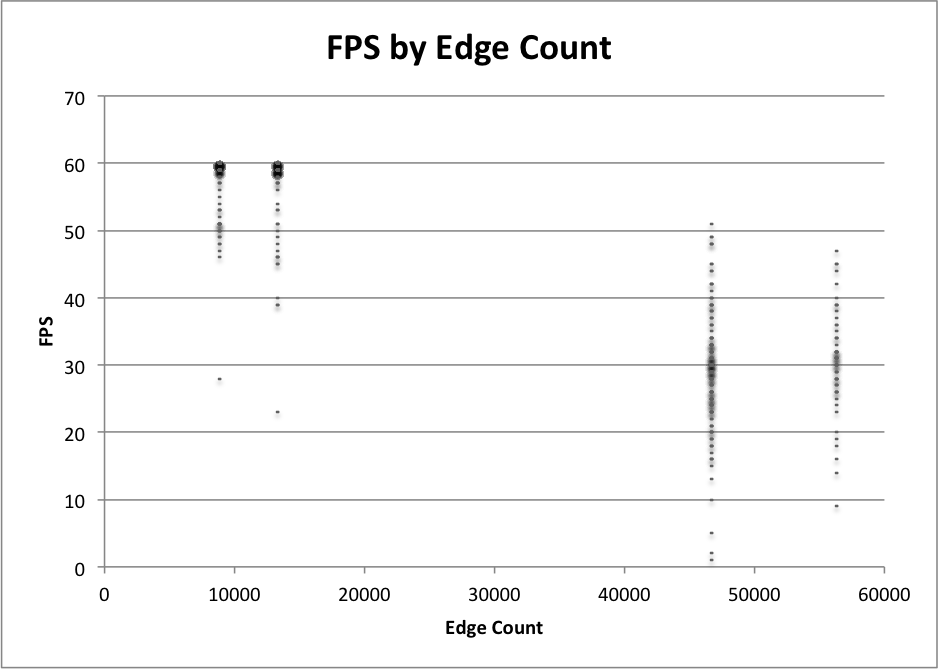}
     \caption{A graph of the phone client's frame rate against a few graphs with different edge counts. Up to a certain point, the FPS remains relatively stable at around 60FPS. However, after a certain point, it fluctuates by a significant amount and drops to an average of about 30FPS.}
     \label{fig:my_label}
 \end{figure}
However, there was still room for improvement. Due to the low rendering power of the Samsung Note 4 phones, we found that loading large graphs could be too cumbersome for the graphics processing unit. Graphs with large amount of nodes and edges could produce significant graphical lag. We found that edges contributed more to this lag, as it required more pixels to render. Figure \ref{fig:my_label} shows sampled frames per second (FPS) counts of a few different sample graphs. We can see that larger graphs created a sudden drop-off in FPS. While we were able to avoid major graphical lag by loading graphs with fewer nodes and edges, we would ideally like to find ways to push that limit. Different shading models and other rendering techniques could assist here. Additionally, we also aim to explore optimizations such as foveated rendering and re-sampling of the graph to maintain the graph representation in a meaningful way while lowering the number of visible nodes.

In addition to the graphical optimizations listed above, we would also like to expand upon the interaction and gesture library we have created. While our work was sufficient to demonstrate the effectiveness of gesture controls in a virtual environment for data visualization, we believe that a broader toolset would be ideal for taking advantage of our system. Ultimately, we would like to see a system such as this contain all the tools necessary for data visualization and analysis, ranging from computation interfaces to graph manipulation tools.

Finally, it is important to note that virtual reality is currently a constantly evolving technology. As technology improves, we would like to adapt this work to adhere to the most lightweight system possible, as that was the goal when choosing the GearVR and Perception Neuron for this project. The OptiTrack system, while providing us with the possibility to do wireless tracking, would ideally be replaced with a cheaper tracking system that could be more financially available.

\section{Conclusion}
We have demonstrated a system that allows users to generate and collaboratively inspect large network layouts, using their hands in a way that is familiar and intuitive. We hope that new data visualization modalities like GraphiteVR will help make complex structures like social networks seem more familiar and intuitive as well.

\acknowledgments{}

\bibliographystyle{abbrv}
\nocite{*}

\bibliography{template}
\end{document}